\documentclass[12pt]{article}
\usepackage{amstext}
\usepackage{amsmath}
\usepackage{amssymb}
\usepackage{relsize}
\usepackage{graphicx}
\usepackage{subfigure}
\usepackage{multirow}
\usepackage[round, longnamesfirst]{natbib}
\usepackage[latin1]{inputenc}
\usepackage{fullpage}
\newcommand{\mc}[1]{\mathcal{#1}}
\title{Tobit Bayesian Model Averaging and the Determinants of Foreign Direct Investment}

\author{Alexander Jordan and Alex Lenkoski\footnote{\noindent \textit{Corresponding author address:} Alex Lenkoski, Institute of Applied Mathematics, Heidelberg University, Im Neuenheimer Feld 294, 69120 Heidelberg, Germany
\newline{E-mail: alex.lenkoski@uni-heidelberg.de}}\\
\textit{Heidelberg University, Germany}
}

\begin{document}
\maketitle
\linespread{1}
\begin{abstract}
We develop a fully Bayesian, computationally efficient framework for incorporating model uncertainty into Type II Tobit models and apply this to the investigation of the determinants of Foreign Direct Investment (FDI).  While direct evaluation of model probabilities is intractable in this setting, we show that by using conditional Bayes Factors, which nest model moves inside a Gibbs sampler, we are able to incorporate model uncertainty in a straight-forward fashion.  We conclude with a study of global FDI flows between 1988-2000.

\end{abstract}
\linespread{1.3}
\newpage
\section{Introduction}
\indent We consider modeling the determinants of Foreign Direct Investment (FDI) and incorporating model uncertainty in this framework.  FDI flows are important for several reasons. For one, they can be taken as a measure of growing economic globalization. Every time a firm sets up a new subsidiary in a different country they engage in FDI. And every positive FDI flow increases the FDI stock in a certain country. The size of the FDI stock shows how much this country is integrated into the international market, because FDI usually comes with a considerable amount of trade as one third of world trade is intrafirm \citep{navven:1996}. Also, FDI may have a positive effect on the host country's economy. Countries may hope for increasing employment and other positive effects like technological spill-over, educational effects on the labor force and of course increasing tax income. While traditionally most FDI occurs between the developed countries, developing countries have growing FDI inflows and this trend is not expected to change.\\
\indent We consider bilateral FDI inflows from the years 1988-2000 and model the decision to engage in foreign direct investment which includes the decisions of how much to invest and whether to invest at all. The two-stage nature of FDI decisions disqualifies the use of simple linear regression and requires a statistical approach that accounts for sample selection.  One framework that may be used is the generalized Tobit model, a two-equation linear regression model.\\
\indent An additional statistical complication arises from the complexity of the FDI decision process.  This leads to a profusion of economic theories and concomitant model uncertainty which we address through Bayesian Model Averaging (BMA).  The specifics of Tobit estimation in a Bayesian context make direct model comparison difficult, due to the inability to resolve nested sets of integrals.  In a pervious study \citep{eich2011} of FDI determinants, this issue was addressed through the use of BIC approximations in a framework based on Heckit estimation.  While the method was useful at illuminating important FDI determinants, it stood on tenuous ground theoretically.  The variable inclusion probabilities were approximations and the entire modeling framework was not strictly Bayesian.  We resolve these issues in the developments below.\\
\indent We show that by nesting model moves inside a larger Gibbs sampler, we are able to compute conditional Bayes Factors (CBFs) directly and easily.  This leads to a procedure that simultaneously addresses selection bias and model uncertainty and does so with limited additional computational complications.  CBFs were originally introduced by \citet{dickey_gunel_1978} and recently proposed in a related econometric setting by \citet{karl_lenkoski_2012}.  Our new method, which we call Tobit Bayesian Model Averaging (TBMA), thus combines CBFs with Markov Chain Monte Carlo (MCMC) estimation to yield a fully theoretically valid methodology that is also computationally much more efficient than the procedure developed in \citet{eich2011}.\\
\indent We conclude with an analysis of the data provided in \citet{eich2011}.  We show that our new method yields broadly similar conclusions as that of \citet{eich2011}, in the determinants of the amount of FDI.  However, the two methods differ substantially in the selection equation.  This is attributed to the fact that the HeckitBMA method of \citet{eich2011} did not ``feed back'' information from the outcome equation into the model probabilities of the selection equation.  Our method, by contrast, updates the two model spaces in a joint fashion, which is more grounded in Bayesian theory. Futher, the method requires only a fraction of the computing time taken by \citet{eich2011}, due to the efficiency of the CBF calculations.\\
\indent The article is organized as follows.  Section~\ref{sec:tobit} will review the Type II Tobit model and discuss the Bayesian estimation of its parameters.  Section~\ref{sec:tbma} will build on these developments to introduce the TBMA algorithm that incorporates model uncertainty.  Section~\ref{sec:fdi} will comprise our main data analysis of FDI determinants.  In Section~\ref{sec:conclude} we conclude.
\section{Review of the Tobit model}\label{sec:tobit}
FDI flow data contain a large number of 0's, resulting from an issue of sample selection \citep{eich2011}.  In order to avoid selection bias we use the generalized Tobit model, also known as Type II Tobit model or sample selection model. The bias in the OLS estimate arises when manually selecting observations (i.e. only use observations with a positive FDI flow). This problem can be solved by using a system of two equations with a selection and an outcome equation thus putting the selection of observations into a probabilistic framework.
The generalized Tobit model for $i \in \{1,2,...,n\}$ is given by
\begin{align}
z_i &=\boldsymbol{w}_i^{'}\boldsymbol{\theta} + \epsilon_i\label{eq:selection}\\
y^{*}_i &=\boldsymbol{x}_i^{'}\boldsymbol{\beta} + \eta_i\label{eq:outcome}\\
y_i &=y_i^{*}\mathbf{1}(z_i \geq 0).\nonumber
\end{align}
Above, $y_i$ is the observed dependent variable and $y^*_i$ its uncensored factor.  $z_i$ is a latent variable that determines whether $y_i$ is observed and $\boldsymbol{w}_i$, $\boldsymbol{x}_i$ are independent $p\times 1$ and $q\times 1$ variable vectors with corresponding coefficient vectors $\boldsymbol{\theta}$, $\boldsymbol{\beta}$. Relating to the modeling of foreign direct investment, $y_i$ is the log FDI flow from one country to another (only one direction) and $\boldsymbol{x}_i$ contains the covariates which are believed to influence the magnitude of FDI flows. The vector $\boldsymbol{w}_i$ consists of the covariates that influence the initial decision whether to engage in FDI.\\
\indent It is reasonable to assume that these two vectors share some covariates. Corruption, for example, could deter firms because of ethical reasons or fear of legal punishment, while the higher cost of doing business could decrease the amount of investment. But the covariate vectors could also be entirely different.\\
\indent We assume correlation between the selection and outcome process and therefore the error term is jointly distributed
$(\epsilon_i, \eta_i)^{'}\sim_{\text{i.i.d.}} \mathcal{N} (\boldsymbol{0},\boldsymbol{\Sigma})$ with covariance matrix
\begin{equation}
\boldsymbol{\Sigma}= \begin{pmatrix} 1 & \gamma \\ \gamma & \phi + \gamma^2 \label{eq:covar}\end{pmatrix}.
\end{equation}
The entry $\Sigma_{11}$ can be set to $1$ for identification because $z_i$ is an unobserved latent variable and only its sign is of relevance \citep{mcculloch2000}.
\subsection{Bayesian Posterior Determination}
\indent In order to determine the posterior distribution of the parameters in (\ref{eq:selection})-(\ref{eq:covar}) we follow the Gibbs sampler outlined in \citet{omori2007}.  This returns a dependent sample whose distribution resembles that of $\boldsymbol{\theta},\boldsymbol{\beta},\gamma,\phi|\boldsymbol{z},\boldsymbol{y}$ where $\boldsymbol{z}=(z_1,...,z_n)^{'}$ and $\boldsymbol{y}=(y_1,...,y_n)^{'}$. \\
\indent As $\boldsymbol{z}$ is a latent variable and cannot be observed, it will be generated using data augmentation \citep{albchib1993}. Early Bayesian approaches to Tobit estimation also imputed the censored observations in $\boldsymbol{y}$. This however is inefficient \citep{chib2007} and \citet{omori2007} outlines a Gibbs sampler based on the marginalization of the posterior distribution over censored observations.\\
\indent Let $\boldsymbol{y}_o$ denote the vector of uncensored observations. We define the following prior distributions
\begin{equation}
\boldsymbol{\theta}\sim\mathcal{N}(\boldsymbol{\theta}_0, \boldsymbol{\Theta}_0)\quad
\boldsymbol{\beta}\sim\mathcal{N}(\boldsymbol{\beta}_0,\boldsymbol{B}_0)\quad
\gamma \sim \mathcal{N}(\gamma_0, G_0)\quad
\phi \sim \mathcal{IG}\left(\tfrac{s_0}{2},\tfrac{S_0}{2}\right),
\end{equation}
where $\mathcal{IG}$ denotes the inverse gamma distribution. This yields the following joint posterior probability density
\begin{equation*}
\text{pr}(\boldsymbol{z},\boldsymbol{\theta},\boldsymbol{\beta},\gamma,\phi|\boldsymbol{y}_o)\propto\text{pr}(\boldsymbol{z},\boldsymbol{y}_o|\boldsymbol{\theta},\boldsymbol{\beta},\gamma,\phi)\ \text{pr}(\boldsymbol{\theta})\ \text{pr}(\boldsymbol{\beta})\ \text{pr}(\gamma)\ \text{pr}(\phi),\hspace{40mm}
\end{equation*}
where
\begin{equation*}
\begin{split}
&\text{pr}(\boldsymbol{z},\boldsymbol{y}_o|\boldsymbol{\theta},\boldsymbol{\beta},\gamma,\phi)\\ &\propto \phi^{-n_o/2}\ \text{exp}\Bigg(-\frac{1}{2}\Bigg[\sum_{i| z_i<0}{(z_i-\boldsymbol{w}_i^{'}\boldsymbol{\theta})^2}\\
&\hspace{27mm}+\sum_{i| z_i\geq0}{(1+\tfrac{\gamma^2}{\phi})(z_i-\boldsymbol{w}_i^{'}\boldsymbol{\theta})^2 - 2\tfrac{\gamma}{\phi}(z_i-\boldsymbol{w}_i^{'}\boldsymbol{\theta})(y_i-\boldsymbol{x}_i^{'}\boldsymbol{\beta}) + \tfrac{1}{\phi}(y_i-\boldsymbol{x}_i^{'}\boldsymbol{\beta})^2}\Bigg]\Bigg)
\end{split}
\end{equation*}
and $n_o$ is the number of uncensored observations.
The conditional posterior distributions of $\boldsymbol{\psi}=(\boldsymbol{\theta}^{'},\boldsymbol{\beta}^{'})^{'}$, $\gamma$ and $\phi$ are then given by
\begin{equation}
\begin{split}
\boldsymbol{\psi}|\boldsymbol{z},\boldsymbol{y_o},\gamma,\phi&\sim\mathcal{N}(\boldsymbol{\psi}_1,\boldsymbol{\Psi}_1) \\
\gamma|\boldsymbol{z},\boldsymbol{y_o},\boldsymbol{\psi},\phi&\sim\mathcal{N}(\gamma_1,G_1) \\
\phi|\boldsymbol{z},\boldsymbol{y_o},\boldsymbol{\psi},\gamma&\sim\mathcal{IG}\left(\tfrac{s_1}{2},\tfrac{S_1}{2}\right).
\end{split}
\end{equation}
A detailed description of the parameters is given in Appendix A.\\
\indent The conditional distributions used to sample the latent variables $z_i$ have the form of normal distributions truncated at zero. A censored observation corresponds to a negative value of $z_i$, whereas for an uncensored observation to occur it must be positive. Due to marginalization over the censored observations, the way in which mean and variance are calculated depends on whether the observation is censored (see Appendix A).
\section{Incorporating Model Uncertainty}\label{sec:tbma}
In the case of FDI, model uncertainty means uncertainty about which FDI determinants have an influence on the actual FDI flows and thus which economic theories can be supported. The two equations in the Tobit model represent different types of influence. Inclusion of a determinant in the selection equation means it has an influence on the decision whether to invest at all, while inclusion in the outcome equation suggests influence on the magnitude of the actual FDI flow.\\
\indent In order to explain the motivation behind our CBF approach, we first review some basic results from classic BMA literature.  We then show how the concept of Bayes Factors can be usefully embedded in a Gibbs sampler yielding CBFs.  These CBFs are then shown to yield straightforward calculations.  The section concludes with an overview of the full TBMA procedure.
\subsection{Bayes Factors}
In a general framework, incorporating model uncertainty involves considering a collection of candidate models $\mc{I}$, using the data $\mathcal{D}$. Each model $I$ consists of a collection of probability distributions for the data $\mathcal{D}$, $\{pr(\mathcal{D}|\xi),\xi \in \Xi_{I}\}$ where $\Xi_{I}$ denotes the parameter space for the parameters of model $I$ and is a subset of the full parameter space $\Xi$.\\
\indent By letting the model become an additional parameter to be assessed in the posterior, we aim to calculate the posterior model probabilities given the data $\mathcal{D}$. By Bayes' rule 
\begin{equation}\label{pmp}
 pr(I|\mathcal{D})=\frac{pr(\mathcal{D}|I)pr(I)}{\sum_{I'\in\mc{I}}{pr(\mathcal{D}|I')pr(I')}},
 \end{equation}
where $pr(I)$, denotes the prior probability for model $I\in\mc{I}$.\\
\indent The integrated likelihood $pr(\mathcal{D}|I)$, is defined by
\begin{equation}\label{marglik}
 pr(\mathcal{D}|I)=\int_{\Xi_I} pr(\mathcal{D}|\xi)pr(\xi|I) d\xi,
 \end{equation}
where $pr(\xi|I)$ is the prior for $\xi$ under model $I$, which by definition has all its mass on $\Xi_I$.\\ 
\indent One possibility for pairwise comparison of models is offered by the Bayes factor (BF), which is in most cases defined together with the posterior odds \citep{kass_raftery_1995}.  The posterior odds of model $I$ versus model $I'$ is given by
\begin{equation*} 
\frac{pr(I|\mathcal{D})}{pr(I'|\mathcal{D})}=\frac{pr(\mathcal{D}|I)}{pr(\mathcal{D}|I')}\frac{pr(I)}{pr(I')}, 
\end{equation*} where 
\begin{equation*} 
\frac{pr(\mathcal{D}|I)}{pr(\mathcal{D}|I')} \quad\text{and}\quad \frac{pr(I)}{pr(I')}
\end{equation*} 
denote the Bayes factor and the prior odds of $I$ versus $I'$, respectively.\\
\indent When the integrated likelihood (\ref{marglik}) and thus the BF can be computed directly, a straightforward method for exploring the model space, Markov Chain Monte Carlo Model Composition (MC3), was developed by \citet{madigan_york_1995}.\\
\indent MC3 determines posterior model probabilities by generating a stochastic process that moves through the model space $\mathcal{I}$ and has equilibrium distribution $pr(I|\mathcal{D})$. Given the current state $I^{(s)}$, MC3 proposes a new model $I'$ according to a proposal distribution $q(\cdot|\cdot)$, calculates
$$
\alpha = \frac{pr(\mathcal{D}|I')pr(I')q(I^{(s)}|I')}{pr(\mathcal{D}|I^{(s)})pr(I^{(s)})q(I'|I^{(s)})}
$$
and sets $I^{(s + 1)} = I'$ with probability $\min\{\alpha,1\}$ otherwise setting $I^{(s + 1)} = I^{(s)}$.
\subsection{Conditional Bayes Factors}
For the model space of the Tobit model we will be using the Cartesian product of two model spaces, one for each equation. Let $\mathcal{I}$ be the model space for the selection equation and $\mathcal{L}$ the model space for the outcome equation. Each model $I_i\in \mathcal{I}$ and $L_j\in\mathcal{L}$ defines which covariates shall be used in their respective equation and the entirety of models $\mathcal{M} = \mc{I}\times \mc{L}$ covers all possible combinations of covariates for both equations. A model $M\in\mc{M}$ can be interpreted as a specific selection of restrictions $\Psi_M$ on the vector $\mathbf{\psi}$. The entry in the parameter vector will be set to $0$ if the corresponding covariate should not be included in the estimation according to the model in question.\\
\indent As discussed above, two competing models could be compared by computing
$$
pr(\mc{D}|M) = \int_{\Sigma}\int_{\psi\in\Psi_{M}} pr(\mc{D}|\psi, \Sigma)pr(\psi|M)pr(\Sigma)d\psi d\Sigma
$$
Unfortunately, this factor, has no obvious, analytic form.  Instead, we focus on the conditional Bayes Factor (CBF),
\begin{equation*}
\text{pr}({D}|M_{ij},\Sigma) = \int \text{pr}({D}|M_{ij},\psi,\Sigma)\ \text{pr}(\psi|M_{ij},\Sigma)d\psi,
\end{equation*}
\indent The CBF thus calculates the integrated likelihood conditional on a fixed setting of $\Sigma$.  In appendix B we show that
\begin{equation}
\text{pr}(\mc{D}|M,\Sigma)\propto\frac{|\boldsymbol{\Psi}_{1,M}|^{1/2}}{|\boldsymbol{\Psi}_{0,M}|^{1/2}}\frac{\text{exp}(-\tfrac{1}{2}\boldsymbol{\psi}_{0,M}^{'}\boldsymbol{\Psi}_{0,M}^{-1}\boldsymbol{\psi}_{0,M})}{\text{exp}(-\tfrac{1}{2}\boldsymbol{\psi}_{1,M}^{'}\boldsymbol{\Psi}_{1,M}^{-1}\boldsymbol{\psi}_{1,M})},\label{eq:cbf}
\end{equation}
where $\psi_{0,M},\psi_{1,M},\Psi_{0,M}$ and $\Psi_{1,M}$ are exactly analogous to the paramters discussed in Section~\ref{sec:tobit} but restricted to the subset of variables contained in model $M$.  We see that the expression (\ref{eq:cbf}) is quite straightforward to calculate and essentially only requires the parameters necessary in the Gibbs sampler.\\
\indent We feel a comment on the implication of equation~(\ref{eq:cbf}) is in order. Remember that for linear regression BIC is calculated as
\begin{equation*}
\text{BIC}_{M} = n\ \text{log}(1-R_{M}^2)+p_{M}\ \text{log}\ n
\end{equation*}
where $n$ is the sample size, $R_{M}^2$ is the value of $R^2$ for the model $M$ and $p_{M}$ is the number of parameters included \citep{raftery_1995}.
Assuming a flat prior distribution on the model space we can use BIC to approximate the posterior model probability
\begin{equation*}
\text{pr}(M|D)\approx\text{exp}(-\tfrac{1}{2}\ \text{BIC}_{M})\approx\frac{1}{n^{p_{M}/2}}\text{exp}(\tfrac{n}{2}R_{M}^2)
\end{equation*}
where $c$ is a normalizing constant. The last part of the equation can be interpreted as the product of a term penalizing for model complexity and a term measuring the goodness of fit. We note that (\ref{eq:cbf}) can be interpreted in the same manner. The second part of the product in (\ref{eq:cbf}) can be rewritten as
\begin{equation*}
\begin{split}
\frac{\text{exp}(-\tfrac{1}{2}\boldsymbol{\psi}_{0,ij}^{'}\boldsymbol{\Psi}_{0,ij}^{-1}\boldsymbol{\psi}_{0,ij})}{\text{exp}(-\tfrac{1}{2}\boldsymbol{\psi}_{1,ij}^{'}\boldsymbol{\Psi}_{1,ij}^{-1}\boldsymbol{\psi}_{1,ij})}\propto\qquad& \\
\text{exp}\Bigg(-\tfrac{1}{2}\bigg[&\sum_{n|z_n<0}{(\boldsymbol{\tilde{y}}_n-\boldsymbol{\tilde{X}}_n \boldsymbol{\psi}_{1,ij})'(\boldsymbol{\tilde{y}}_n-\boldsymbol{\tilde{X}}_n \boldsymbol{\psi}_{1,ij})}\\
+&\sum_{n|z_n\geq 0}{(\boldsymbol{\tilde{y}}_n-\boldsymbol{\tilde{X}}_n \boldsymbol{\psi}_{1,ij})'\Sigma^{-1}(\boldsymbol{\tilde{y}}_n-\boldsymbol{\tilde{X}}_n \boldsymbol{\psi}_{1,ij})}\\
+&\ \  (\boldsymbol{\psi}_{0,ij}-\boldsymbol{\psi}_{1,ij})'\boldsymbol{\Psi}_{0,ij}^{-1}(\boldsymbol{\psi}_{0,ij}-\boldsymbol{\psi}_{1,ij})\bigg]\Bigg),
\end{split}
\end{equation*}
where $\boldsymbol{\tilde{y}}_n$ is a two-dimensional vector of the latent and dependent variable and $\boldsymbol{\tilde{X}}_n$ is a $2\times(p+q)$ matrix of covariates (see Appendix A for the definitions). Essentially, this is the residual sum of squares for the decorelated and normalized equations of the Tobit model as the argument of a strictly decreasing function. A decrease in the residual sum of squares, or increase in goodness of fit, leads to a greater integrated likelihood. The ratio containing the prior and posterior covariance matrix penalizes for increasing model complexity.  We therefore see that (\ref{eq:cbf}) has a strong relationship to the calculation of BIC, and yet relies on no approximations.
\subsection{Tobit Bayesian Model Averaging (TBMA)}
\indent We now review the entire TBMA algorithm that integrates MC3 into the Gibbs sampler using CBFs. We assume a flat prior on the model space (i.e. $pr(M) \propto 1$), although other prior distributions could be readily accommodated. For a detailed description of the parameters see Appendix A.
\begin{enumerate}
\item{Initialize $\boldsymbol{\psi}$, $\gamma$ and $\phi$.}
\item{Sample $\boldsymbol{z}|\boldsymbol{y}_o,\boldsymbol{\psi},\gamma,\phi$.
  \begin{itemize}
  \item{For censored observations: $\boldsymbol{z}_i|M,\boldsymbol{\psi},\gamma,\phi\sim\mathcal{N}_{(-\infty,0)}(\boldsymbol{w}_i^{'}\boldsymbol{\theta},1)$.}
  \item{For uncensored observations: $\boldsymbol{z}_i|y_i,M,\boldsymbol{\psi},\gamma,\phi\sim\mathcal{N}_{\left[0,\infty\right)}(\mu_i,\sigma_i^2)$.}
  \end{itemize}
}
\item{Sample $\gamma|M,\boldsymbol{z},\boldsymbol{y_o},\boldsymbol{\psi},\phi\sim\mathcal{N}(\gamma_1,G_1)$.}
\item{Sample $\phi|M,\boldsymbol{z},\boldsymbol{y_o},\boldsymbol{\psi},\gamma\sim\mathcal{IG}\left(\tfrac{s_1}{2},\tfrac{S_1}{2}\right)$.}
\item{Apply modified MC3.
  \begin{enumerate}
  \item{Sample $M'$ from the neighborhood of $M$.}
  \item{Compute pr$(\mc{D}|M',\gamma,\phi)$ and pr$(\mc{D}|M,\gamma,\phi)$ as in equation (\ref{eq:cbf}).}
  \item{Accept $M'$ as new $M$ with probability $\text{min} \left\{1,\text{CBF}_{M'M}\right\}$.}
  \end{enumerate}
}
\item{Sample $\boldsymbol{\psi}|M,\boldsymbol{z},\boldsymbol{y_o},\gamma,\phi\sim\mathcal{N}(\boldsymbol{\psi}_{1},\boldsymbol{\Psi}_{1})$.}
\item{Go to 2.}
\end{enumerate}
By adding one single step to the existing Gibbs sampler we are able to include model uncertainty. Furthermore, it is a very efficient way since the calculations for the parameters of the conditional posterior distribution of $\boldsymbol{\psi}$ are included in step 5.
\section{Application to FDI Data}\label{sec:fdi}
\subsection{Economic theories and associated FDI determinants}
Every investor considering FDI is confronted with a two-stage decision. Whether he should make an investment abroad and if so, how much he should invest. Even though later on we will be looking at aggregate FDI flows derived from the balances of payments of different countries, these decisions are made at the firm level. Naturally, firm-level data would be the best basis for inference but they are almost impossible to come by.\\
\indent Probably the most important principle in the theoretic work on FDI is the OLI framework by \citet{dunn1980}. The OLI principle states that firms decide to engage in FDI if they have \textit{ownership} of goods or production processes that give them market power, if they can benefit from having a plant in a foreign \textit{location} rather than at home and if \textit{internalization} of the foreign activity provides an advantage over contracts with local firms. Ownership and location-specific advantages are reasons why firms would want to produce in a foreign country in the first place, and internalization is a way to protect your assets in face of market failures. Every time assets are revealed there is a risk of dissipation which would result in a loss of value. This concern may lead to difficulties in negotiations for rents of a certain asset. Firms would not want to reveal the full asset before a contract is finalized while the potential local partner is not ready to pay the full price until the asset has been fully revealed. Even after a contract is finalized, there is the risk of hold-up situations when the local partner had to make investments in relationship-specific goods. Due to incomplete contracts, there is a possibility of renegotiations where the local partner has little bargaining power. These negotiation difficulties can make a firm internalize their production by engaging in FDI rather than signing contracts with local firms.\\
\indent Early analytical FDI theory based on the OLI principle suggested two distinct motivations for FDI, horizontal and vertical FDI \citep{mark1984, help1984}. Horizontal FDI (HFDI) is undertaken when firms want to access different markets. Instead of increasing their production at home and exporting produced goods, they invest in production in the foreign country. It is a way to serve a foreign market by partially duplicating the existing production. Thus, a similar development level in both host and source country for the industry in which the firm is operating may increase HFDI. Reasons to engage in HFDI in the first place include the evasion of trade restrictions and reduction of trade costs which can be summarized as tariff-jumping FDI. This lowers the marginal costs for the firm and may lead to strategic advantage over national firms in the foreign market. Vertical FDI (VFDI), on the other hand, leverages lower factor prices in other economies. For example, a firm may have an interest to move unskilled labour-intensive activities to countries with low wages when the trade costs are relatively low as well. Usually firms with easily segmentable production processes tend to engage in VFDI. These two motivations have been unified into the knowledge-capital model of FDI \citep{mark1997}. Determinants suggested by these theories are trade costs, educational differences and market size. Note that aggregated FDI flows are currently still dominated by horizontal FDI which is closely related to the fact that most FDI flows take place between developed countries.\\
Developed countries also often engage in trade or investment agreements, which are designed to facilitate these international transactions. The question related to these agreements is whether they do have the positive impact they are supposed to have and in the case of trade agreements how they affect FDI. Liberalization of trade may have a negative effect when trade and FDI are substitutes, as in HFDI theory, but trading can lead to a higher GDP, which means a greater market size, and that is generally thought to have a positive impact on FDI. Overall, trade agreements are expected to positively influence FDI since a large portion of world trade is intra-firm. Another means of facilitation of both trade and investment are currency unions and these as such should not be expected to have a negative impact.\\
\indent Tax treaties and taxation in general, intuitively should have a negative impact on FDI. However, the empirical results are ambiguous. Taxation reduces the wealth of a firm and reduces the possibilities of investment but the laws vary strongly between different countries. Firms may be motivated to evade taxes by strategic investment, which could increase investment and some countries may try to attract FDI with tax incentives. Tax treaties are a way for countries to level the playing field, resulting in higher overall tax.\\
\indent An approach, that is very successful and popular in empirical analysis of FDI, uses gravity equations. Gravity equations have been used for almost half a century to explain ex post effects of various country characteristics on bilateral trade flows \citep{berg2007}. In physics, we can observe that the gravitational force is poportional to the size of the physical body and inversely proportional to the square of the distance from the physical body. This leads to the idea that, as in physics, a greater market size of both host and source country positively influences trade flows while distance between the two economies has a negative effect. Only recently, a theoretical foundation for the gravity model in trade flows has been established \citep{andvwin2003, berg2007}. Since trade decisions are closely related to FDI decisions and FDI may be considered a substitute for trade in HFDI theory, the gravity model is a very popular approach in examining FDI determinants \citep{navven:1996}. However, it seems as though FDI behavior is much more complicated to model than trade flows and there is no paper that identifies gravity variables as the sole determinants of FDI flows \citep{blon2005}.\\
\indent One reason why FDI flows are more difficult to model than trade flows could be because the location of investment is a part of the decision. When national firms are looking to invest with the intention to export or import, the decision is influenced by potential trade partners and whether the conditions in the home country encourage further investment. The actual trade decision is not very sensitive to country characteristics, except those that directly affect profit. Those may be trade costs that affect the profit margin and market size which has an influence on the total demand of goods.
National firms have to consider only the risks they take when investing in their home country, whereas multinational firms have to evaluate their disintegration costs which include all their risks in several different countries. They try to find the best country for their investment and many country characteristics have a potential influence on the risk an investor takes when making a long-term investment. These range from cumbersome regulations to economic or political conditions which in turn may lead to early withdrawal or even total loss of investment. While accurately measuring the risks is very difficult, estimation of the effects on FDI should be easier. Since individual analysis of all countries is too costly even for multinational firms, decisions are often influenced by indices that try to measure political, legal and economic conditions. The effect of indices or ratings on financial decisions can clearly be seen in the current debt crisis in Europe. For FDI, legal protection of assets, corruption and general quality of institutions but also existing or potential conflicts are only some of the factors measured by indices. If in a certain country, firms have to fear expropriation or difficulties in the repatriation of profits one should expect a lower FDI inflow for said country. Corruption increases cost of doing business and in some cases may be heavily punished (U.S. Foreign Corrupt Practices Act) while weak institutions may be pressured in face of a change of government. Internal and external conflicts may lead to unforeseen structural changes and falls in profitability. Internal conflicts can range from civil disorder to civil war caused by religious/ethnic tensions or socioeconomic factors like unemployment or poverty. External conflicts can present themselves in foreign pressures, like restrictions on operations or trade/investment sanctions, to full-out war.
\subsection{Description of the Dataset}
We use the same set of data as \citet{eich2011} which is an unbalanced panel that covers the years 1988 - 2000. It contains 803 unique country pairs of 46 countries. 21 of these countries are not members of the OECD. The total number of observations is 14863 and 64 percent of these observations indicate zero FDI inflow. In addition to the observed FDI inflows it contains data on 55 potential FDI determinants. The descriptive statistics of the log FDI flows and all 55 determinants can be found in Table 1.\\
\indent The dataset is based on \citet{rst2008} which is the source for the gravity variables (market size, proxied by the real GDP, and distance), for factor endowments (development levels, proxied by real GDP per capita, and educational difference), and various other determinants - namely, productivity (real GDP per worker), GDP growth rate, financial risk and common language. Also, we will be using a variable indicating past FDI flows. Razin et al. obtained the FDI outflow data from the OECD International Direct Investment Database (OECD) which is noted in U.S. dollars and deflated it by the U.S. Consumer Price Index. Market potential has been constructed according to the definition given in \citet{blon2007}.\\
\indent \citet{eichhenn2011} provided the geographical and historical indicators, border and colony, but also a list of both multilateral and bilateral trade agreements, as well as currency union indicators. Another list of bilateral tax treaties has been obtained from \citet{neuspess2005}.\\
\indent The average effective corporate tax rates have been constructed using the definitions and information provided in \citet{alt1988}, \citet{blondav2004} and the U.S. Treasury Corporate Tax Files.\\
Institutional variables and other risk factors have been taken from the International Country Risk Guides \citep{icrg}, where the exact definition of each index can be found. Included in the dataset are indices of host and source country's corruption, bureaucratic efficiency and investment profile, a composite index concerned with contract viability/expropriation, repatriation of profits and payment delays, but also several political risk indices. These indices proxy the benefits of democratic accountability, government stability and a strong legal system, but also the risks of military government participation, internal/external conflicts, ethnic/religious tension and the socioeconomic profile, which includes unemployment and poverty. 

\linespread{1}
\begin{table}[ht]
\caption{Descriptive Statistics}
\begin{center}
\scalebox{0.7}{
\begin{tabular}{lrrrr|l}
 & mean & sd & min & max & \textbf{Source} \\ 
  \hline
APECij & 0.09 & 0.29 & 0 & 1 & Eicher, Henn (2011) \\ 
  BI\_RTAij & 0.01 & 0.11 & 0 & 1 & Eicher, Henn (2011) \\ 
  BORDERij & 0.04 & 0.19 & 0 & 1 & Eicher, Henn (2011) \\ 
  BUREAUi & 3.15 & 0.95 & 0 & 4 & International Country Risk Guide \\ 
  BUREAUj & 3.18 & 0.95 & 0 & 4 & International Country Risk Guide \\ 
  COLONYij & 0.03 & 0.18 & 0 & 1 & Eicher, Henn (2011) \\ 
  COM\_LANGij & 0.18 & 0.38 & 0 & 1 & Razin, Sadka, Tong (2008) \\ 
  CORRUPTi & 4.23 & 1.32 & 1.08 & 6 & International Country Risk Guide \\ 
  CORRUPTj & 4.25 & 1.33 & 1.08 & 6 & International Country Risk Guide \\ 
  DEMOCRATICi & 4.9 & 1.25 & 1 & 6 & International Country Risk Guide \\ 
  DEMOCRATICj & 4.96 & 1.21 & 1 & 6 & International Country Risk Guide \\ 
  DEVELOPMENTi & 5.38 & 1.37 & 1.71 & 9.1 & constructed from Razin, Sadka, Tong (2008) \\ 
  DEVELOPMENTj & 5.32 & 1.37 & 1.71 & 9.1 & constructed from Razin, Sadka, Tong (2008) \\ 
  DISTANCEij & 8.24 & 0.92 & 4.92 & 9.42 & Razin, Sadka, Tong (2008) \\ 
  DOLLARij & 0 & 0.04 & 0 & 1 & Eicher, Henn (2011) \\ 
  EDU\_DIFFij & -0.06 & 3.22 & -8.5 & 9.89 & Razin, Sadka, Tong (2008) \\ 
  EEAij & 0.08 & 0.26 & 0 & 1 & Eicher, Henn (2011) \\ 
  EFTAij & 0.01 & 0.09 & 0 & 1 & Eicher, Henn (2011) \\ 
  ETHNIC\_TENSIONi & 4.82 & 1.27 & 1 & 6 & International Country Risk Guide \\ 
  ETHNIC\_TENSIONj & 4.81 & 1.33 & 1 & 6 & International Country Risk Guide \\ 
  EUij & 0.1 & 0.29 & 0 & 1 & Eicher, Henn (2011) \\ 
  EUROij & 0.01 & 0.1 & 0 & 1 & Eicher, Henn (2011) \\ 
  EXTERN\_CONFLICTi & 10.88 & 1.5 & 4.25 & 12 & International Country Risk Guide \\ 
  EXTERN\_CONFLICTj & 10.86 & 1.59 & 4.25 & 12 & International Country Risk Guide \\ 
  FIN\_RISKi & 39.83 & 7.28 & 18 & 50 & Razin, Sadka, Tong (2008) \\ 
  FIN\_RISKj & 39.83 & 7.38 & 18 & 50 & Razin, Sadka, Tong (2008) \\ 
  GDP\_GROWTHi & 0.04 & 0.04 & -0.13 & 0.14 & constructed from Razin, Sadka, Tong (2008) \\ 
  GDP\_GROWTHj & 0.04 & 0.05 & -0.13 & 0.45 & constructed from Razin, Sadka, Tong (2008) \\ 
  GOV\_STABILITYi & 7.61 & 2.02 & 1 & 12 & International Country Risk Guide \\ 
  GOV\_STABILITYj & 7.57 & 2.04 & 1 & 11 & International Country Risk Guide \\ 
  INTERN\_CONFLICTi & 10.06 & 2.21 & 3 & 12 & International Country Risk Guide \\ 
  INTERN\_CONFLICTj & 10.02 & 2.28 & 3 & 12 & International Country Risk Guide \\ 
  INV\_PROFi & 6.97 & 1.73 & 2.33 & 11.17 & International Country Risk Guide \\ 
  INV\_PROFj & 6.97 & 1.74 & 2.42 & 11.17 & International Country Risk Guide \\ 
  INVEST\_TREATYij & 0.12 & 0.32 & 0 & 1 & Neumayer, Spess (2005) \\ 
  LAIAij & 0.02 & 0.15 & 0 & 1 & Eicher, Henn (2011) \\ 
  LAW\_ORDERi & 4.71 & 1.4 & 1 & 6 & International Country Risk Guide \\ 
  LAW\_ORDERj & 4.71 & 1.44 & 1 & 6 & International Country Risk Guide \\ 
  LOG\_FDI & 1.27 & 2.26 & -2.85 & 11.14 & Razin, Sadka, Tong (2008) \\ 
  MILITARYi & 4.82 & 1.5 & 1 & 6 & International Country Risk Guide \\ 
  MILITARYj & 4.85 & 1.52 & 0 & 6 & International Country Risk Guide \\ 
  MRKT\_POTENTIALj & 0.57 & 0.2 & 0.34 & 1.42 & constructed see Blonigen et al., 2007) \\ 
  MRKT\_SIZEi & 9.17 & 1.17 & 5.81 & 10.75 & Razin, Sadka, Tong (2008) \\ 
  MRKT\_SIZEj & 9.24 & 1.1 & 6.06 & 10.75 & Razin, Sadka, Tong (2008) \\ 
  NAFTAij & 0 & 0.06 & 0 & 1 & Eicher, Henn (2011) \\ 
  NEG\_FDI\_LAG & 0.05 & 0.22 & 0 & 1 & constructed from Razin, Sadka, Tong (2008) \\ 
  PAST\_FDI & 0.34 & 0.47 & 0 & 1 & Razin, Sadka, Tong (2008) \\ 
  PRODUCTIVITYi & 36.32 & 18.44 & 2.67 & 74.66 & Razin, Sadka, Tong (2008) \\ 
  PRODUCTIVITYj & 37.25 & 18 & 4.24 & 74.66 & Razin, Sadka, Tong (2008) \\ 
  RELIGIOUS\_TENSIONi & 5.2 & 1.07 & 1 & 6 & International Country Risk Guide \\ 
  RELIGIOUS\_TENSIONj & 5.14 & 1.16 & 1 & 6 & International Country Risk Guide \\ 
  RERij & 103.51 & 31.55 & 16.73 & 597.64 & USDA http://www.ers.usda.gov \\ 
  SOCIO\_ECONi & 6.66 & 1.64 & 2 & 11 & International Country Risk Guide \\ 
  SOCIO\_ECONj & 6.68 & 1.65 & 2 & 11 & International Country Risk Guide \\ 
  TAXi & 0.22 & 0.11 & 0 & 0.73 & 1980-92: Altshulter et al. (1998) \\ 
  TAXj & 0.23 & 0.11 & 0 & 0.73 & 1994-02: IRS/SOI, World Tax Database \\ 
   \hline
\end{tabular}
}
\end{center}
\end{table}
\linespread{1.5}

\linespread{1}
\begin{table}[ht]
\caption{FDI Determinants and their Estimated Effects in the Past}
\begin{center}
\scalebox{0.67}{
\begin{tabular}{cl|ccc|l}
 \multirow{2}{*}{} & \multirow{2}{*}{Variable Name} & \multicolumn{3}{|c|}{Estimated Effect} & \multirow{2}{*}{Variable Description}\\
 & & + & none & - & \\ \hline
 \multirow{3}{*}{Gravity} & DISTANCEij & & 1 & 16 & natural log of bilateral distance\\
 & MRKT\_SIZEi & 8 & 2 & & source natural log of real GDP\\
 & MRKT\_SIZEj & 13 & 5 & 2 & host natural log of real GDP\\ \hline
 \multirow{3}{*}{Geography/History} & BORDERij & 2 & 3 & & =1 if pair share a common border\\
 & COLONYij & 4 & 2 & & =1 if pair share colonial relationship\\
 & COM\_LANGij & 10 & 3 & & =1 if pair share common language\\ \hline
 \multirow{3} {*} {Factor Endowment} & DEVELOPMENTi & 3 & 4 & & source natural log of real GDP per capita\\
 & DEVELOPMENTj & 7 & 7 & & host natural log of real GDP per capita\\
 & EDU\_DIFFij & 2 & 4 & 2 & source minus host education level\\ \hline
 \multirow{5} {*} {Growth \& Productivity} & GDP\_GROWTHi & & & & source GDP growth rate\\
 & GDP\_GROWTHj & 2 & 3 & & host GDP growth rate\\
 & MRKT\_POTENTIALj & 1 & 1 & & $\begin{matrix} \text{sum of host's distance-weighted GDP}\\\text{to all other countries} \end{matrix}$\\
 & PRODUCTIVITYi & 1 & 1 & 1 & source productivity (real GDP per worker)\\
 & PRODUCTIVITYj & 1 & 1 & & host productivity (real GDP per worker)\\ \hline
 \multirow{3}{*} {Fiscal / Monetary Policy} & TAXi & & 1 & & source effective corporate tax rate\\
 & TAXj & & 3 & 5 & host effective corporate tax rate\\
 & RERij & & 4 & 2 & real exchange rate (host/source currency)\\ \hline
 \multirow{11}{*}{$\begin{matrix} \text{RTAs / CUs /}\\ \text{Investment} \end{matrix}$}& INVEST\_TREATYij & 1 & 3 & & \multirow{11}{*}{=1 if both countries are in a treaty} \\
 & RTAij & 0 & 0 & 0 & \\
 & BI\_RTAij & 1 & 3 & 1 & \\
 & NAFTAij & 1 & 3 & 1 & \\
 & EUij & 1 & 3 & & \\
 & EFTAij & 1 & 1 & & \\
 & EEAij & & & & \\
 & LAIAij & & & & \\
 & APECij & 1 & 2 & & \\
 & EUROij & & & & \\
 & DOLLARij & & & & \\ \hline
 \multirow{6}{*}{Economic Risk}& BUREAUi & & & & source bureaucratic quality\\
 & BUREAUj & 2 & & & host bureaucratic quality\\
 & CORRUPTi & & & & source corruption\\
 & CORRUPTj & 3 & 2 & & host corruption\\
 & FIN\_RISKi & 1 & 2 & 2 & source financial risk\\
 & FIN\_RISKj & 2 & 2 & & host financial risk\\ \hline
 \multirow{20}{*}{Political Risk} & DEMOCRATICi & & & & source democratic accountability\\
 & DEMOCRATICj & 1 & & & host democratic accountability\\
 & ETHNIC\_TENSIONi & & & & source ethnic tensions\\
 & ETHNIC\_TENSIONj & & 1 & & host ethnic tensions\\
 & EXTERN\_CONFLICTi & & & & source external conflict\\
 & EXTERN\_CONFLICTj & 1 & & & host external conflict\\
 & GOV\_STABILITYi & & & & source government stability\\
 & GOV\_STABILITYj & 2 & & & host government stability\\
 & INTERN\_CONFLICTi & & & & source internal conflict\\
 & INTERN\_CONFLICTj & & 1 & & host internal conflict\\
 & INV\_PROFILEi & & & & source investment profile\\
 & INV\_PROFILEj & 2 & & & host investment profile\\
 & LAW\_ORDERi & & 1 & & source law and order\\
 & LAW\_ORDERj & 2 & 1 & & host law and order\\
 & MILITARYi & & & & source military in politics\\
 & MILITARYj & & 1 & & host military in politics\\
 & RELIGIOUS\_TENSIONi & & & & source religious tensions\\
 & RELIGIOUS\_TENSIONj & & 1 & & host religious tensions\\
 & SOCIO\_ECONi & & & & source socioeconomic conditions\\
 & SOCIO\_ECONj & & 1 & & host socioeconomic conditions\\ \hline
 \multicolumn{6}{l}{}\\
 \multicolumn{6}{l}{source: \citet{eich2011}}
\end{tabular}
}
\end{center}
\end{table}
\linespread{1.5}
\subsection{Results}
We ran several chains with 100,000 iterations each, to confirm convergence in the model and parameter space. As we can see in Figure 1, the running average number of determinants suggested by TBMA converges relatively quickly and is consistent with different starting points, for both the selection and the outcome equation. The average percentage of model jumps is just above 7\%. Using TBMA we obtain 19 determinants for the selection equation and 28 for the outcome equation with an inclusion probability greather than 50\%. This is more than the numbers suggested by HeckitBMA of 13 and 23 but still less than the 23 and 35 determinants of the Heckit model \citep{eich2011}.

\linespread{1}
\begin{table}[ht]
\caption{TBMA estimates for FDI determinants}
\begin{center}
\scalebox{0.7}{
\begin{tabular}{l|rrr|rrr}
&\multicolumn{3}{|c|}{\textbf{FDI selection}}&\multicolumn{3}{c}{\textbf{FDI flow}}\\
 & incl.prob & post.mean & post.sdev & incl.prob & post.mean & post.sdev \\ 
  \hline
APECij & 0.19 & 0.14 & 0.06 & 1.00 & 0.87 & 0.09 \\ 
  COLONYij & \textbf{1.00} & 0.44 & 0.10 & 1.00 & 1.11 & 0.11 \\ 
  COM\_LANGij & 0.12 & 0.09 & 0.05 & 1.00 & 0.62 & 0.07 \\ 
  CORRUPTi & \textbf{1.00} & 0.10 & 0.02 & 1.00 & 0.26 & 0.04 \\ 
  CORRUPTj & \textbf{0.59} & 0.07 & 0.02 & 1.00 & 0.17 & 0.03 \\ 
  DEVELOPMENTi & \textbf{1.00} & 0.28 & 0.02 & 1.00 & 0.94 & 0.03 \\ 
  DEVELOPMENTj & \textbf{1.00} & 0.28 & 0.02 & 1.00 & 0.80 & 0.03 \\ 
  DISTANCEij & \textbf{1.00} & -0.32 & 0.03 & 1.00 & -0.76 & 0.03 \\ 
  DOLLARij & 0.31 & -0.56 & 0.35 & 1.00 & 4.00 & 0.71 \\ 
  ETHNIC\_TENSIONi & \textbf{0.61} & 0.05 & 0.02 & 1.00 & 0.17 & 0.03 \\ 
  INTERN\_CONFLICTi & 0.00 & -0.01 & 0.01 & 1.00 & -0.16 & 0.02 \\ 
  MRKT\_POTENTIALj & 0.37 & 0.21 & 0.09 & 1.00 & -0.92 & 0.12 \\ 
  MRKT\_SIZEi & 0.41 & 0.10 & 0.03 & 1.00 & 0.39 & 0.08 \\ 
  MRKT\_SIZEj & \textbf{1.00} & -0.34 & 0.06 & 1.00 & -1.30 & 0.07 \\ 
  PAST\_FDI & \textbf{1.00} & 2.24 & 0.04 & 1.00 & 1.11 & 0.14 \\ 
  PRODUCTIVITYj & \textbf{0.71} & 0.01 & 0.00 & 1.00 & 0.04 & 0.00 \\ 
  RELIGIOUS\_TENSIONj & \textbf{0.95} & 0.09 & 0.02 & 1.00 & 0.38 & 0.03 \\ 
  SOCIO\_ECONj & 0.01 & 0.01 & 0.01 & 1.00 & 0.08 & 0.02 \\ 
  TAXi & \textbf{0.81} & -0.55 & 0.18 & 1.00 & -3.42 & 0.26 \\ 
  TAXj & \textbf{0.99} & -0.84 & 0.18 & 1.00 & -4.36 & 0.26 \\ 
  BUREAUi & 0.02 & 0.05 & 0.03 & 1.00 & 0.33 & 0.07 \\ 
  SOCIO\_ECONi & \textbf{0.78} & 0.05 & 0.01 & 1.00 & 0.12 & 0.02 \\ 
  GDP\_GROWTHj & 0.11 & 0.29 & 0.39 & 0.96 & 2.01 & 0.64 \\ 
  INTERN\_CONFLICTj & 0.02 & 0.02 & 0.02 & 0.92 & 0.08 & 0.02 \\ 
  LAIAij & 0.46 & -0.34 & 0.14 & 0.90 & -0.89 & 0.29 \\ 
  EXTERN\_CONFLICTj & 0.03 & -0.03 & 0.01 & 0.86 & -0.08 & 0.02 \\ 
  EEAij & 0.05 & -0.07 & 0.09 & 0.82 & -0.26 & 0.08 \\ 
  PRODUCTIVITYi & 0.00 & 0.00 & 0.00 & 0.78 & 0.01 & 0.00 \\ 
  BI\_RTAij & 0.44 & 0.32 & 0.13 & 0.33 & 0.37 & 0.19 \\ 
  GDP\_GROWTHi & 0.39 & -0.85 & 0.56 & 0.32 & -0.85 & 0.89 \\ 
  DEMOCRATICi & 0.01 & 0.02 & 0.02 & 0.26 & 0.10 & 0.04 \\ 
  MILITARYj & 0.14 & 0.06 & 0.02 & 0.26 & 0.11 & 0.04 \\ 
  NEG\_FDI\_LAG & \textbf{1.00} & 0.82 & 0.08 & 0.20 & -0.16 & 0.07 \\ 
  NAFTAij & 0.09 & 0.09 & 0.36 & 0.12 & 0.27 & 0.26 \\ 
  BUREAUj & 0.03 & 0.06 & 0.04 & 0.11 & 0.12 & 0.06 \\ 
  EFTAij & 0.34 & 0.39 & 0.19 & 0.08 & -0.17 & 0.17 \\ 
  LAW\_ORDERi & 0.03 & 0.04 & 0.02 & 0.06 & -0.07 & 0.04 \\ 
  EUROij & 0.05 & -0.02 & 0.17 & 0.06 & -0.06 & 0.18 \\ 
  INVEST\_TREATYij & 0.06 & 0.09 & 0.05 & 0.06 & 0.10 & 0.08 \\ 
  INV\_PROFi & 0.01 & -0.02 & 0.02 & 0.05 & 0.04 & 0.02 \\ 
  EUij & 0.32 & -0.17 & 0.07 & 0.05 & 0.04 & 0.11 \\ 
  BORDERij & \textbf{1.00} & -0.73 & 0.11 & 0.04 & 0.04 & 0.12 \\ 
  INV\_PROFj & 0.00 & -0.01 & 0.01 & 0.02 & -0.03 & 0.02 \\ 
  MILITARYi & \textbf{0.64} & 0.08 & 0.02 & 0.02 & 0.02 & 0.04 \\ 
  EXTERN\_CONFLICTi & \textbf{1.00} & -0.08 & 0.01 & 0.02 & -0.03 & 0.02 \\ 
  LAW\_ORDERj & \textbf{0.52} & 0.08 & 0.02 & 0.02 & 0.01 & 0.04 \\ 
  DEMOCRATICj & 0.00 & -0.01 & 0.02 & 0.01 & 0.04 & 0.04 \\ 
  RERij & 0.00 & -0.00 & 0.00 & 0.01 & -0.00 & 0.00 \\ 
  ETHNIC\_TENSIONj & 0.09 & 0.04 & 0.02 & 0.01 & 0.02 & 0.03 \\ 
  GOV\_STABILITYi & 0.00 & -0.02 & 0.01 & 0.01 & 0.01 & 0.02 \\ 
  RELIGIOUS\_TENSIONi & 0.02 & -0.03 & 0.02 & 0.01 & -0.01 & 0.04 \\ 
  GOV\_STABILITYj & 0.00 & -0.00 & 0.01 & 0.01 & 0.01 & 0.01 \\ 
  EDU\_DIFFij & 0.01 & -0.01 & 0.01 & 0.00 & -0.00 & 0.01 \\ 
  FIN\_RISKj & 0.00 & 0.00 & 0.00 & 0.00 & 0.00 & 0.00 \\ 
  FIN\_RISKi & 0.01 & 0.01 & 0.00 & 0.00 & 0.00 & 0.01 \\ 
   \hline
\end{tabular}
}
\end{center}
\end{table}
\linespread{1.5}

\begin{figure}[h]
\subfigure[selection equation]{\includegraphics[width=0.49\textwidth]{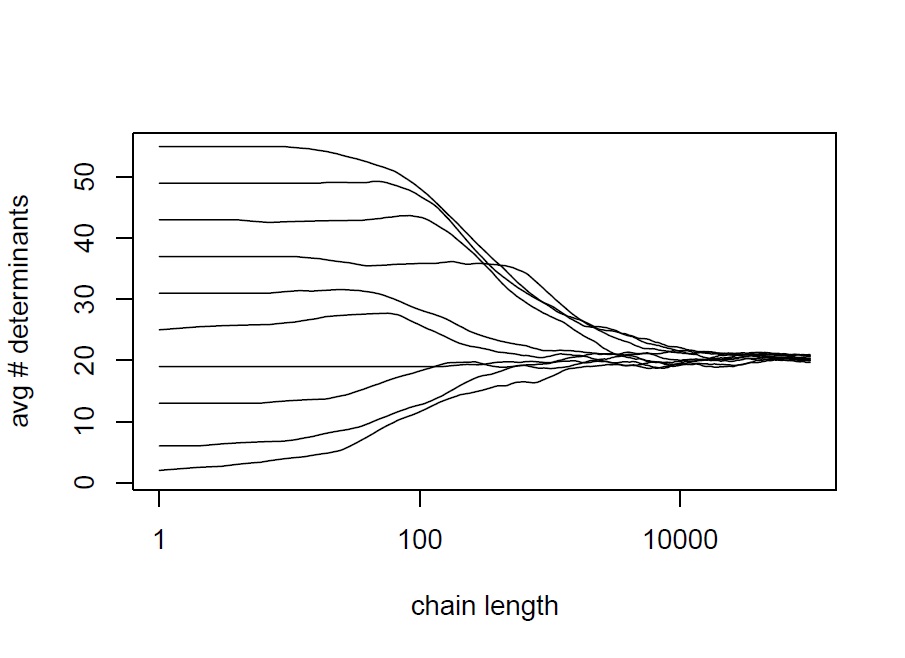}}\hfill
\subfigure[outcome equation]{\includegraphics[width=0.49\textwidth]{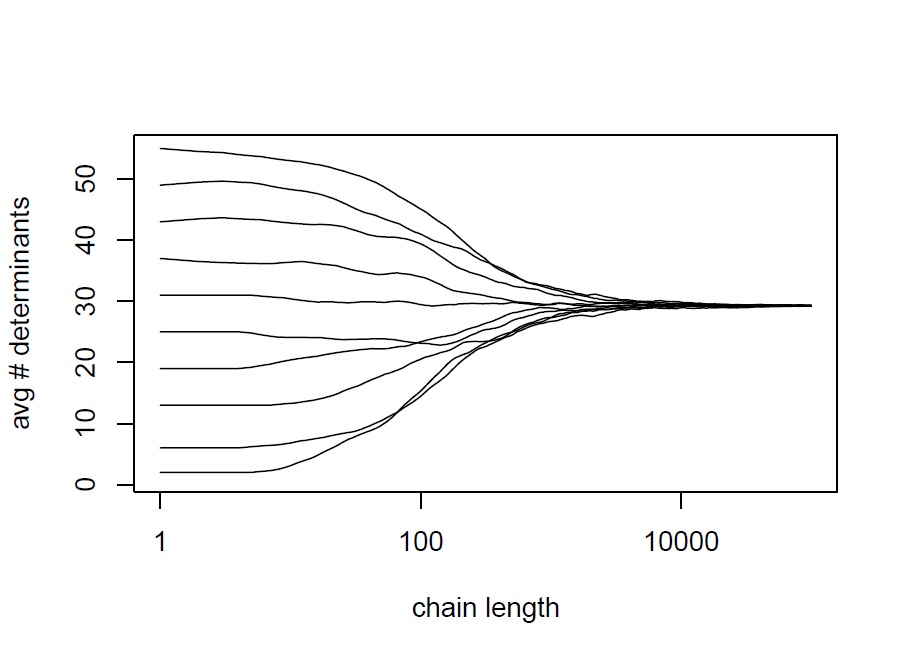}}
\caption{Running average numbers of determinants for selection and outcome equation of 2 chains with length 100k.}
\end{figure}

\indent Table 2 of \citep{eich2011} provides an overview of whether a determinant has been expected to have a positive, negative or no effect in the past. Table 3 shows TBMA estimates of posterior mean and standard deviation, as well as the inclusion probabilities of each determinant. The inclusion probability is defined as the ratio of the number of samples to the length of the chain after discarding the first 10,000 iterations as burn-in period.\\
\indent We find that TBMA gives support for roughly the same determinants as HeckitBMA when modeling the magnitude of FDI flows. As expected, the gravity variables host and source market size as well as distance have a decisive effect on FDI flows. Surprisingly, while distance has a negative and the source country's market size has a positive effect, we find that the market size of the host country influences FDI flows negatively. Of the geographical and historical determinants only a former colonial connection and common language exert a decisive positive effect on the amount of investment. As with HeckitBMA, the levels of development are found to be of significance while an effect of educational difference cannot be supported. A higher effective tax rate of both host and source country negatively influences FDI. GDP growth in the host country, as well as productivity in the host country are very likely to have a positive impact, whereas source country GDP growth and productivity do not seem to be of importance. We also find the same peculiar behavior, that market potential does seem to have an effect although it is a negative one. The effect of regional trade agreements, namely APEC and LAIA are supported by both HeckitBMA and TBMA but only TBMA suggests a significant effect of the EEA. Both methods find a positive impact if the same currency of two countries is the U.S. dollar whereas the EURO seems to be of little importance, keeping in mind that the examined years range from 1988 to 2000. Risk proxies supported by both HeckitBMA and TBMA are corruption in host and source country, religious tension and internal conflicts in the host country and bureaucratic quality and ethnic tension in the source country. HeckitBMA additionally suggests the investment profile of the source country while TBMA shows an effect of the socioeconomic conditions in host and source country and external conflicts in the host country. All risk proxies from the International Country Risk Guides are positively oriented, which means we are talking about lack of tensions and conflicts or quality of conditions.\\
\indent Looking at the selection equation this unison of HeckitBMA and TBMA seems to fade. Only five determinants receive high inclusion probabilities from both methods. These determinants are the levels of development and religious tension in the host country, the effective tax rates in both host and source country and a previous FDI relation. HeckitBMA further suggests the regional trade agreements APEC and EFTA, common language, financial risk in host and source country, religious tension in the source country as well as military government participation and ethnic tension in the host country as determinants with significant effect on the decision whether to enage in FDI. Additional to the five determinants supported by both methods, TBMA suggests that a former colonial relationship, distance, a common border, host market size, host and source development levels and a few risk indices are important for the decision of whether to engage in FDI. These risk indices proxy mainly source country risks, namely corruption, ethnic tension, the socioeconomic conditions, military government participation and external conflicts in the source country and only one host country risk, strength of the legal system and the obeyance thereof.\\
\indent These differences in the inclusion probabilities for the selection equation result from the different ways of estimation. HeckitBMA, as a two-stage estimation method, isolates the estimations for selection and outcome equation and models the influence of the selection on the outcome equation with the Inverse Mills Ratio. In TBMA these two estimations are simultaneous and feedback from the outcome to the selection equation is expected.
\section{Conclusions}\label{sec:conclude}
Dealing with model uncertainty by averaging parameter estimates according to posterior model probabilities has seen a rise in the last few years. MCMC methods are a very efficient way\footnote{The whole algorithm, as implemented in section 4.1, takes less than 1 hour of runtime for our dataset and a chain of length 100k on an Intel(R) Core(TM) i5-2400 CPU @ 3.1GHz, 8GB RAM, Windows 7 SP1 64 Bit system using R x64 2.13.1 (no multicore support).} to approximate the posterior model probabilities and achieve Bayesian Model Averaging, as models with high posterior model probabilities are evaluated more frequently than others with low probability. The length of the sum for the averaged parameter estimate is determined by the length of the Markov chain and not by the total number of models, which is particularly useful if we have a huge number of models where many of them have very low probability. While the necessary length of the chain is dependent on the number of high probability models, MCMC methods help finding these models without the necessity to calculate the posterior probability for every possible model.\\
\indent Within the MCMC framework, we found a way to implement Bayesian Model Averaging for the generalized Tobit model without using approximations, like the BIC, to achieve the right mixture of models according to the Bayes factors. Maximizing the likelihood and then penalizing for model complexity under the assumption of infinite observations, as BIC does, should be avoided especially when data is rare. Within a Gibbs sampler that explores the parameter space, it is possible to use the MC3 method with conditional Bayes factors. This way, estimates of the posterior model probabilities are unbiased for any number of observations, whereas BIC may systematically penalize model complexity too strongly or too weakly when dealing with a small dataset. While the structure of the conditional likelihood for the conditional Bayes factor is similar in interpretation to the structure of the posterior model probability derived from the BIC, it is not necessary to use a flat prior on the model space for BMA. We can easily modify step 5 in the Tobit Bayesian Model Averaging algorithm (section 4.1) to include various prior beliefs about model probabilities, according to section 3.4, equation (8).\\
\indent Since TBMA is fully justified within the Bayesian framework, we expect more accurate parameter estimates in comparison to other BMA results. Also, other hierarchical components, like random effects and time dependencies, can be included more confidently with a firm ground to rely on. TBMA easily accomodates additional components, as all it takes is another additional Gibbs step, and using the combination of conditional Bayes factors and MC3 fits perfectly into the MCMC paradigm.

\newpage
\renewcommand{\theequation}{A-\arabic{equation}} 
\setcounter{equation}{0}  
\section*{Appendix A}
Parameters for the conditional posterior distribution of $z_i$:
\begin{itemize}
\item{
For censored observations:
\begin{equation*}
\begin{split}
&\mu_i = \boldsymbol{w}_i^{'}\boldsymbol{\theta}\hspace{37mm}\\
&\sigma_i^2 = 1
\end{split}
\end{equation*}
}
\item{
For uncensored observations:
\begin{equation*}
\begin{split}
&\mu_i = \boldsymbol{w}_i^{'}\boldsymbol{\theta}+\frac{\gamma}{\phi + \gamma^2}(y_i-\boldsymbol{x}_i^{'}\boldsymbol{\beta})\\
&\sigma_i^2 = 1 - \frac{\gamma^2}{\phi+\gamma^2}
\end{split}
\end{equation*}
}
\end{itemize}
Parameters for the conditional posterior distribution of $\boldsymbol{\psi}$:
\begin{equation*}
\begin{split}
&\boldsymbol{\psi}_1 = \boldsymbol{\Psi}_1 \left(\boldsymbol{\Psi}_0^{-1}\boldsymbol{\psi}_0 + \sum_{i| z_i\geq0} {\boldsymbol{\tilde{X}}_i^{'}\Sigma^{-1}\boldsymbol{\tilde{y}}_i}+\sum_{i| z_i<0} {\boldsymbol{\tilde{X}}_i^{'}\boldsymbol{\tilde{y}}_i}\right) \\
&\boldsymbol{\Psi}_1 = \left(\boldsymbol{\Psi}_0^{-1}+\sum_{i| z_i\geq0} {\boldsymbol{\tilde{X}}_i^{'}\Sigma^{-1}\boldsymbol{\tilde{X}}_i}+\sum_{i| z_i<0} {\boldsymbol{\tilde{X}}_i^{'}\boldsymbol{\tilde{X}}_i}\right)^{-1}\\
&\boldsymbol{\tilde{y}}_i = \begin{cases} \begin{pmatrix} z_i\\y_i \end{pmatrix} & \text{if } z_i\geq0 \\ \begin{pmatrix} z_i\\0 \end{pmatrix} & \text{if } z_i<0 \end{cases}, \qquad \boldsymbol{\tilde{X}}_i = \begin{cases} \begin{pmatrix} \boldsymbol{w}_i^{'} & \boldsymbol{0}^{'}\\ \boldsymbol{0}^{'} & \boldsymbol{x}_i^{'} \end{pmatrix} & \text{if } z_i\geq0 \\ \begin{pmatrix} \boldsymbol{w}_i^{'} & \boldsymbol{0}^{'}\\ \boldsymbol{0}^{'} & \boldsymbol{0}^{'} \end{pmatrix} & \text{if } z_i<0 \end{cases}\\
&\boldsymbol{\psi}_0 = \begin{pmatrix} \boldsymbol{\theta}_0\\ \boldsymbol{\beta}_0 \end{pmatrix},\qquad \boldsymbol{\Psi}_0 = \begin{pmatrix} \boldsymbol{\Theta}_0 & \boldsymbol{0}\\ \boldsymbol{0} & \boldsymbol{B}_0 \end{pmatrix}
\end{split}
\end{equation*}
Parameters for the conditional posterior distribution of $\gamma$:
\begin{equation*}
\begin{split}
&\gamma_1 = G_1\left(G_0^{-1}\gamma_0+\phi^{-1}\sum_{i| z_i\geq0}{(z_i-\boldsymbol{w}_i^{'}\boldsymbol{\theta})(y_i-\boldsymbol{x}_i^{'}\boldsymbol{\beta})}\right)\\
&G_1 = \left(G_0^{-1} + \phi^{-1}\sum_{i| z_i\geq0}{(z_i-\boldsymbol{w}_i^{'}\boldsymbol{\theta})^2}\right)^{-1}
\end{split}
\end{equation*}
Parameters for the conditional posterior distribution of $\phi$:
\begin{equation*}
\begin{split}
&s_1 = s_0 + n_o\\
&S_1 = S_0 + \gamma^2\sum_{i| z_i\geq0}{(z_i-\boldsymbol{w}_i^{'}\boldsymbol{\theta})^2}-2\gamma\sum_{i| z_i\geq0}{(z_i-\boldsymbol{w}_i^{'}\boldsymbol{\theta})(y_i-\boldsymbol{x}_i^{'}\boldsymbol{\beta})}+\sum_{i| z_i\geq0}{(y_i-\boldsymbol{x}_i^{'}\boldsymbol{\beta})^2}
\end{split}
\end{equation*}
\renewcommand{\theequation}{B-\arabic{equation}} 
\setcounter{equation}{0}  
\section*{Appendix B}
The covariance structure of the linear regression system is given by the covariance matrix $\Sigma$, which is parameterized by $\gamma$ and $\phi$ (see section 2.1) and a priori independent from the choice of model $M_{ij}\in\mathcal{M}$. Restrictions are imposed on the space of the regression parameter vector $\boldsymbol{\psi}=(\boldsymbol{\theta}^{'},\boldsymbol{\beta}^{'})^{'}$ according to the model $M_{ij}$ (see section 3) and the dependence is noted by $\boldsymbol{\psi}_{ij}$. The parameters for the conditional prior normal distribution of $\boldsymbol{\psi}_{ij}$ are mean $\boldsymbol{\psi}_{0,ij}$ and covariance matrix $\boldsymbol{\Psi}_{0,ij}$. For the definition of $\boldsymbol{\tilde{y}}_i$ and $\boldsymbol{\tilde{X}}_i$, as well as the posterior mean $\boldsymbol{\psi}_{1,ij}$ and covariance matrix $\boldsymbol{\Psi}_{1,ij}$ see Appendix A.
\begin{multline*}
\shoveright{\text{pr}(\boldsymbol{D}|M_{ij},\gamma,\phi)=\int \text{pr}(\boldsymbol{D}|M_{ij},\boldsymbol{\psi}_{ij},\gamma,\phi)\ \text{pr}(\boldsymbol{\psi}_{ij}|M_{ij})d\boldsymbol{\psi}_{ij}}
\end{multline*}
\begin{multline*}
\propto(2\pi)^{-p_{ij}/2}\ |\boldsymbol{\Psi}_{0,ij}|^{-1/2}\\
\shoveleft{\quad\qquad\qquad\mathlarger{\int}\text{exp}\Bigg(-\tfrac{1}{2}\bigg[\sum_{k|z_k\geq0}{(\tilde{y}_k-\tilde{X}_k\boldsymbol{\psi}_{ij})'\Sigma^{-1}(\tilde{y}_k-\tilde{X}_k\boldsymbol{\psi}_{ij})}}\\
+\sum_{k|z_k<0}{(\tilde{y}_k-\tilde{X}_k\boldsymbol{\psi}_{ij})'(\tilde{y}_k-\tilde{X}_k\boldsymbol{\psi}_{ij})}\\
+(\boldsymbol{\psi}_{ij}-\boldsymbol{\psi}_{0,ij})'\boldsymbol{\Psi}_{0,ij}^{-1}(\boldsymbol{\psi}_{ij}-\boldsymbol{\psi}_{0,ij})\bigg]\Bigg)\ d\boldsymbol{\psi}_{ij}
\end{multline*}
\begin{multline*}
\propto(2\pi)^{-p_{ij}/2}\ |\boldsymbol{\Psi}_{0,ij}|^{-1/2}\ \text{exp}(-\tfrac{1}{2}\boldsymbol{\psi}_{0,ij}'\boldsymbol{\Psi}_{0,ij}^{-1}\boldsymbol{\psi}_{0,ij})\\
\mathlarger{\int}\text{exp}\Bigg(-\tfrac{1}{2}\bigg[-2\Big(\boldsymbol{\psi}_{0,ij}'\boldsymbol{\Psi}_{0,ij}^{-1}+\sum_{k|z_k\geq0}{\tilde{y}_k'\Sigma^{-1}\tilde{X}_k}+\sum_{k|z_k<0}{\tilde{y}_k'\tilde{X}_k}\Big)\boldsymbol{\psi}_{ij}\\
+\boldsymbol{\psi}_{ij}'\Big(\boldsymbol{\Psi}_0^{-1}+\sum_{k|z_k\geq0}{\tilde{X}_k'\Sigma^{-1}\tilde{X}_k}+\sum_{k|z_k<0}{\tilde{X}_k'\tilde{X}_k}\Big)\boldsymbol{\psi}_{ij}\bigg]\Bigg)\ d\boldsymbol{\psi}_{ij}
\end{multline*}
\begin{multline*}
=(2\pi)^{-p_{ij}/2}\ |\boldsymbol{\Psi}_{0,ij}|^{-1/2}\ \text{exp}(-\tfrac{1}{2}\boldsymbol{\psi}_{0,ij}'\boldsymbol{\Psi}_{0,ij}^{-1}\boldsymbol{\psi}_{0,ij})\\
\mathlarger{\int}\text{exp}\Bigg(-\tfrac{1}{2}\bigg[-2\Big(\boldsymbol{\psi}_{0,ij}'\boldsymbol{\Psi}_{0,ij}^{-1}+\sum_{k|z_k\geq0}{\tilde{y}_k'\Sigma^{-1}\tilde{X}_k}+\sum_{k|z_k<0}{\tilde{y}_k'\tilde{X}_k}\Big)\boldsymbol{\Psi}_{1,ij}\boldsymbol{\Psi}_{1,ij}^{-1}\boldsymbol{\psi}_{ij}\\
+\boldsymbol{\psi}_{ij}'\boldsymbol{\Psi}_{1,ij}^{-1}\boldsymbol{\psi}_{ij}\bigg]\Bigg)\ d\boldsymbol{\psi}_{ij}
\end{multline*}
\begin{multline*}
=(2\pi)^{-p_{ij}/2}\ |\boldsymbol{\Psi}_{0,ij}|^{-1/2}\ \text{exp}(-\tfrac{1}{2}\boldsymbol{\psi}_{0,ij}'\boldsymbol{\Psi}_{0,ij}^{-1}\boldsymbol{\psi}_{0,ij})\\
\mathlarger{\int}\text{exp}\bigg(-\tfrac{1}{2}\Big[-2\boldsymbol{\psi}_{1,ij}'\boldsymbol{\Psi}_{1,ij}^{-1}\boldsymbol{\psi}_{ij} + \boldsymbol{\psi}_{ij}'\boldsymbol{\Psi}_{1,ij}^{-1}\boldsymbol{\psi}_{ij}\Big]\bigg)\ d\boldsymbol{\psi}_{ij}\\
\end{multline*}
\begin{multline*}
=(2\pi)^{-p_{ij}/2}\ |\boldsymbol{\Psi}_{1,ij}|^{-1/2}\ |\boldsymbol{\Psi}_{1,ij}|^{1/2}\ |\boldsymbol{\Psi}_{0,ij}|^{-1/2}\ \text{exp}(-\tfrac{1}{2}\boldsymbol{\psi}_{0,ij}'\boldsymbol{\Psi}_{0,ij}^{-1}\boldsymbol{\psi}_{0,ij})\ \text{exp}(\tfrac{1}{2}\boldsymbol{\psi}_{1,ij}'\boldsymbol{\Psi}_{1,ij}^{-1}\boldsymbol{\psi}_{1,ij})\\
\mathlarger{\int}\text{exp}\bigg(-\tfrac{1}{2}\Big[\boldsymbol{\psi}_{1,ij}'\boldsymbol{\Psi}_{1,ij}^{-1}\boldsymbol{\psi}_{1,ij}-2\boldsymbol{\psi}_{1,ij}'\boldsymbol{\Psi}_{1,ij}^{-1}\boldsymbol{\psi}_{ij} + \boldsymbol{\psi}_{ij}'\boldsymbol{\Psi}_{1,ij}^{-1}\boldsymbol{\psi}_{ij}\Big]\bigg)\ d\boldsymbol{\psi}_{ij}\\
\end{multline*}
\begin{multline*}
=|\boldsymbol{\Psi}_{0,ij}|^{-1/2}\ \text{exp}(-\tfrac{1}{2}\boldsymbol{\psi}_{0,ij}'\boldsymbol{\Psi}_{0,ij}^{-1}\boldsymbol{\psi}_{0,ij})\ |\boldsymbol{\Psi}_{1,ij}|^{1/2}\ \text{exp}(\tfrac{1}{2}\boldsymbol{\psi}_{1,ij}'\boldsymbol{\Psi}_{1,ij}^{-1}\boldsymbol{\psi}_{1,ij})\\
\square
\end{multline*}

\nocite{hoeting1999}
\bibliography{JordanLenkoski}
\end{document}